\documentclass[
		journal=cgdefu,
		manuscript=article,
		layout=twocolumn
	]{achemso}
	\setkeys{acs}{etalmode=truncate}
	\setkeys{acs}{maxauthors=3}

\usepackage[version=3]{mhchem}
\usepackage[detect-all,separate-uncertainty=true]{siunitx}
\usepackage[font=footnotesize]{caption}
\usepackage{cuted}

\author{Simon Fischer}
\author{Jon-Olaf Krisponeit}
\email{krisponeit@ifp.uni-bremen.de}
\affiliation[University of Bremen]
{\footnotesize Institute of Solid State Physics, University of Bremen, Otto-Hahn-Allee 1, 28359 Bremen, Germany}
\alsoaffiliation[MAPEX Bremen]
{MAPEX Center for Materials and Processes, University of Bremen, 28359 Bremen, Germany}
\author{Michael Foerster}
\author{Lucia Aballe}
\affiliation[ALBA Synchrotron Barcelona]
{ALBA Synchrotron Light Facility, Carrer de la Llum 2-26, 08290 Cerdanyola del Vall\`es, Barcelona, Spain}
\author{Jens Falta}
\affiliation[University of Bremen]
{\footnotesize Institute of Solid State Physics, University of Bremen, Otto-Hahn-Allee 1, 28359 Bremen, Germany}
\alsoaffiliation[MAPEX Bremen]
{MAPEX Center for Materials and Processes, University of Bremen, 28359 Bremen, Germany}
\author{Jan Ingo Flege}
\affiliation[University of Cottbus-Senftenberg]
{Applied Physics and Semiconductor Spectroscopy, Brandenburg University of Technology Cottbus-Senftenberg, Konrad-Zuse-Str. 1, 03046 Cottbus, Germany}

\title{Massively strained \ce{VO2} thin film growth on \ce{RuO2}}

\abbreviations{XAS,XPEEM,LEEM,XPS,LEED,MBE}

\begin{document}

\begin{strip}
	\centering
	\small
	\begin{minipage}{0.85\textwidth}
	\begin{abstract}
		Strain engineering vanadium dioxide thin films is one way to alter this material's characteristic first order transition from semiconductor to metal. In this study we extend the exploitable strain regime by utilizing the very large lattice mismatch of \SI{8.78}{\percent} occurring in the \ce{VO2}/\ce{RuO2} system along the $c$ axis of the rutile structure.
		
		We have grown \ce{VO2} thin films on single domain \ce{RuO2} islands of two distinct surface orientations by atomic oxygen-supported reactive MBE. These films were examined by spatially resolved photoelectron and x-ray absorption spectroscopy, confirming the correct stoichiometry.
		Low energy electron diffraction then reveals the \ce{VO2} films to grow indeed fully strained on \ce{RuO2}(110), exhibiting a previously unreported (2$\times$2) reconstruction. On \ce{TiO2}(110) substrates, we reproduce this reconstruction and attribute it to an oxygen-rich termination caused by the high oxygen chemical potential. On \ce{RuO2}(100) on the other hand, the films grow fully relaxed.
		Hence, the presented growth method allows for simultaneous access to a remarkable strain window ranging from bulk-like structures to massively strained regions.
	\end{abstract}
	\end{minipage}
\end{strip}

\section{Introduction}
\begin{figure}[bt]
		\centering
		\includegraphics[width=\linewidth]{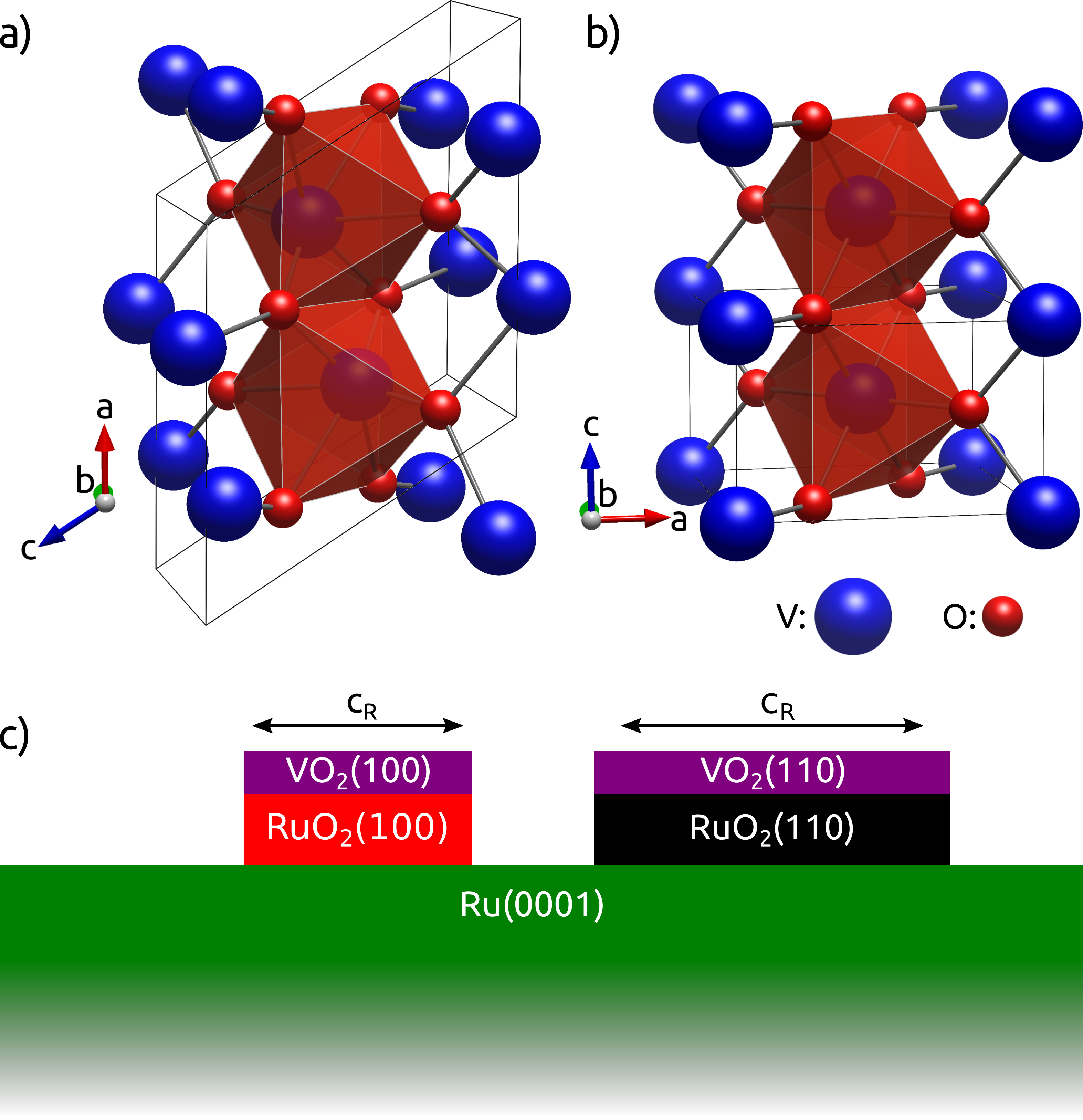}
		\caption{\ce{VO2} structure in the a) monoclinic semiconducting (space group $P2_1/c$) and b) rutile metallic (space group $P4_2/mnm$) phase. The images are created using VESTA 3\cite{momma2011}. Thin black lines represent the unit cell edges. The dimerization and zigzag-like displacement of vanadium atoms in the monoclinic cell is exaggerated. c) Schematic cross section of strained \ce{VO2} films on \ce{RuO2} islands. $c_R$ denotes the $c$ axis of the rutile structure.}
		\label{fig:structure}
	\end{figure}

	Bulk vanadium dioxide exhibits a temperature-induced semiconductor-metal transition at \SI{68}{\celsius} \cite{morin1959}. This change in resistivity is accompanied by a structural transition from a semiconducting monoclinic (M) phase to a metallic phase of rutile (R) structure (see Figure \ref{fig:structure}a and b) where the vanadium atoms along the rutile $c_R$ axis are dimerized in the monoclinic phase with the dimers additionally tilting in a zigzag-like pattern perpendicular to that direction. Because photoemission studies showed that the lattice change alone cannot account for the band gap opening \cite{koethe2006}, this is commonly interpreted as a Peierls-assisted Mott transition \cite{weber2012}.
	
	Applying moderate (epitaxial) stress along the aforementioned $c_R$ axis changes the temperature of the lattice transition and in consequence also the electronic transition. This tunability enables and facilitates applications in switching devices like ``smart'' windows that change their IR reflectivity at a desired temperature \cite{manning2004} or in micro actuators utilizing the structural transition \cite{liu2012}.
	Applying larger stress however may separate the electronic and the lattice-driven transition, making an insulating phase of \emph{rutile} structure possible \cite{laverock2012,abreu2012}. On the other hand, a metallic monoclinic phase can be achieved by doping the \ce{VO2} with electrons \cite{shiga2019}. This charge-injection induced phase switching can also be utilized for field effect devices because it happens on a much faster timescale than the lattice distortion \cite{nakano2012,belyaev2014}.
	
	\begin{table*}[tb]
		\small
		\caption{\ Comparison of lattice parameters for different rutile materials and the mismatch that \ce{VO2} has to the other materials where negative values mean that the \ce{VO2} will be strained compressively}
		\label{tbl:lattice}
		\begin{tabular*}{0.7\textwidth}{@{\extracolsep{\fill}}lccc}
			\hline
			& $c$ (\si{\angstrom})
			& $a$ (\si{\angstrom})
			& $c/a$ \\
			\hline
			rutile \ce{VO2} \cite{rogers1993}
			& 2.8557
			& 4.5540
			& 0.627 \\
			\ce{TiO2}
			& 2.9592 (\SI{3.62}{\percent})
			& 4.5933 (\SI{0.863}{\percent})
			& 0.644 \\
			\ce{RuO2} \cite{rao1969}	
			& 3.1064 (\SI{8.07}{\percent})	
			& 4.4909 (\SI{-1.41}{\percent})	
			& 0.692 \\
			\hline
		\end{tabular*}
	\end{table*}
	
	Previously, strain-tuning \ce{VO2} films has been studied extensively on \ce{TiO2}(001) and \ce{TiO2}(110) substrates \cite{muraoka2002,sambi1997,agnoli2004,quackenbush2015,quackenbush2017,surnev2003} that feature a coincident crystal structure and a comparatively low lattice mismatch. Along the $c_R$ axis, this mismatch amounts to \SI{-0.57}{\percent} for (001)-oriented substrates ($c$ axis out of plane, calculated for the pseudomorphic case using a Poisson's ratio of $\nu=0.249$ \cite{aetukuri2013}) and \SI{3.62}{\percent} for (110)-oriented substrates ($c_R$ axis in plane), respectively. The resulting strain enables transition temperatures of \SI{27}{\celsius} and \SI{96}{\celsius} \cite{muraoka2002}. On \ce{TiO2}(100) substrates, \citet{laverock2012} report a higher actual strain than for \ce{TiO2}(110). This substantial elongation of the $c_R$ axis is increasing electron correlation effects and thus leads to an increasingly Mott-like character of the transition \cite{mukherjee2016, quackenbush2016}.
	
	For devices requiring fast switching, it is desirable to suppress the lattice transition completely. This may be achieved by applying a much higher stress \cite{kittiwatanakul2014,laverock2012}. In order to reach a new regime of lattice mismatch, we consider \ce{RuO2} as the substrate, having a much larger lattice constant especially in the $c_R$ direction (see Table \ref{tbl:lattice}).
	\citet{aetukuri2013} showed that for \ce{VO2} films grown on a \ce{RuO2} buffer layer on a \ce{TiO2}(001) substrate, the transition temperature in principle is tunable between the below-bulk values expected for \ce{TiO2}(001) and a relaxed \ce{RuO2}(001) substrate (\SI{0.92}{\percent} stress in $c$ direction as calculated for the pseudomorphic case using a Poisson's ratio of $\nu=0.249$).
	\citet{cui2012} also report a slightly reduced transition temperature for \ce{VO2} on a thin \ce{RuO2}(101) buffer layer also on \ce{TiO2}.
	In contrast to that, growing \ce{VO2} films on (110)- and (100)-oriented \ce{RuO2} would mean a huge lattice mismatch of \SI{8.78}{\percent} in-plane in the $c_R$ direction. Applying this high level of stress, relaxed growth could be expected. However, in the following we will show that it is possible to grow almost fully strained thin \ce{VO2} films on \ce{RuO2}(110).
	
	In order to simultaneously prepare \ce{VO2} films on different \ce{RuO2} orientations under identical experimental conditions, we use mesoscale \ce{RuO2} islands grown on a Ru(0001) substrate as illustrated in Figure \ref{fig:structure}c. As the distribution of differently oriented \ce{RuO2} islands can be controlled via the Ru crystal oxidation temperature and the oxygen chemical potential \cite{flege2015}, this approach enables us to compare the processes on each orientation simultaneously, as well as on the bare ruthenium, and can also serve as a templating mechanism for future applications. A further advantage is that this growth recipe is available on conductive samples (i.e. Ru crystals) and thus is suited for in-situ studies with real-time low-energy electron-based techniques as there is no significant charging of the surface.
	
	The phase transition quality strongly depends on the exact stoichiometry of the vanadium oxide\cite{griffiths1974}. Especially under near-UHV conditions, oxygen-deficient films may occur. 
	Hence, for this study it is not only important to monitor the local surface structure of the differently oriented substrate phases for determining the amount of \ce{VO2} strain locally, but it is also essential to control the film stoichiometry in situ. This can be achieved by using a SPELEEM\cite{schmidt1998} (spectroscopic photoemission and low energy electron microscope) which combines spatially resolved chemically sensitive methods of photoelectron spectroscopy with the power of local electron diffraction.

\section{Experimental}
	The LEEM and PEEM (low energy electron microscopy~/~photoemission electron microscopy) experiments were performed at the CIRCE beamline of the ALBA Synchrotron, where an Elmitec SPELEEM is in operation\cite{aballe2015}. Additional measurements were performed using the Elmitec LEEM III at the Institute of Solid State Physics in Bremen. Both microscopes have a lateral resolution of about \SI{10}{\nano\meter} in LEEM mode, including intensity-voltage (I(V)) spectroscopy, and down to \SI{20}{\nano\meter} in XPEEM mode (X-ray absorption and X-ray photoemission spectromicroscopy). Low energy electron diffraction patterns can be obtained from regions as small as \SI{250}{\nano\meter}, a technique called \si{\micro}LEED.
	
	The LEEM and PEEM data were analyzed using the program Fiji \cite{schindelin2012}. I(V) curve cluster analysis of the image stacks was performed using the python package scikit-learn \cite{pedregosa2011}: First, the stacks were reduced in dimensionality by principal component analysis, then the K-Means algorithm was applied to sort the individual spectra associated with every image pixel into one of a predefined number of clusters. This allowed for assigning each pixel to a specific surface phase.
	
	As substrate, a commercial Ru(0001) single crystal by Mateck with \SI{< 0.1}{\degree} miscut was used. It was cleaned by repeatedly exposing it to molecular oxygen and then flash-annealing to \SI{>1400}{\celsius} \cite{madey1975}. The substrate then was exposed to atomic oxygen from an OBS 40 thermal cracker (Dr.~Eberl MBE-Komponenten GmbH), which was operated at \SI{1780}{\celsius} with a molecular oxygen background pressure of \SI{2e-6}{\milli\bar}.
	Analogous to the \ce{RuO2} growth recipes using \ce{NO2} and molecular oxygen described in previous studies \cite{flege2009, flege2015}, different \ce{RuO2} orientations were prepared simultaneously. To obtain different distributions of the island orientations, the Ru crystal was oxidized at different sample temperatures.
	
	For the \ce{VO2} film preparation, the reactive molecular beam epitaxy (MBE) method was employed: While providing atomic oxygen from the thermal oxygen cracker, vanadium was evaporated onto the sample using an \mbox{e-beam} evaporator. During growth, the sample temperature was kept at \SI{200}{\celsius} and \SI{400}{\celsius} for different samples. The thermal cracker was again operated at \SI{1780}{\celsius} with a molecular oxygen background pressure of \SI{2e-6}{\milli\bar} while the evaporator was operated at \SI{28}{\watt} heating power applied to the vanadium rod resulting in a deposition rate of about \SI{4}{ML\per\hour}.
	
	As the \ce{VO2} films tend to reduce in UHV when subjected to synchrotron radiation or to electrons with an energy of \SI{> 10}{\electronvolt}, a background pressure of \SI{2e-6}{\milli\bar} molecular \ce{O2} was maintained during the in-situ experiments. Observing a beam-damaged region while heating slowly showed that beam damage suffered in this way can be avoided and repaired, respectively, when annealing at \SI{> 200}{\celsius} in this oxygen-rich environment.
	
	Additional analysis of a characteristic sample was done ex situ at a GIXRD/XRR instrument at the DESY facility in Hamburg. Film thickness and roughness were  analyzed by fitting the XRR spectrum up to $2\theta = \ang{9}$ angle of reflection (see Supporting Information).

\section{Results and discussion}
	
\subsection{Substrate morphology: \ce{RuO2}/Ru orientations and phase distribution}
	\begin{figure}[tb]
		\centering
		\includegraphics[width=\linewidth]{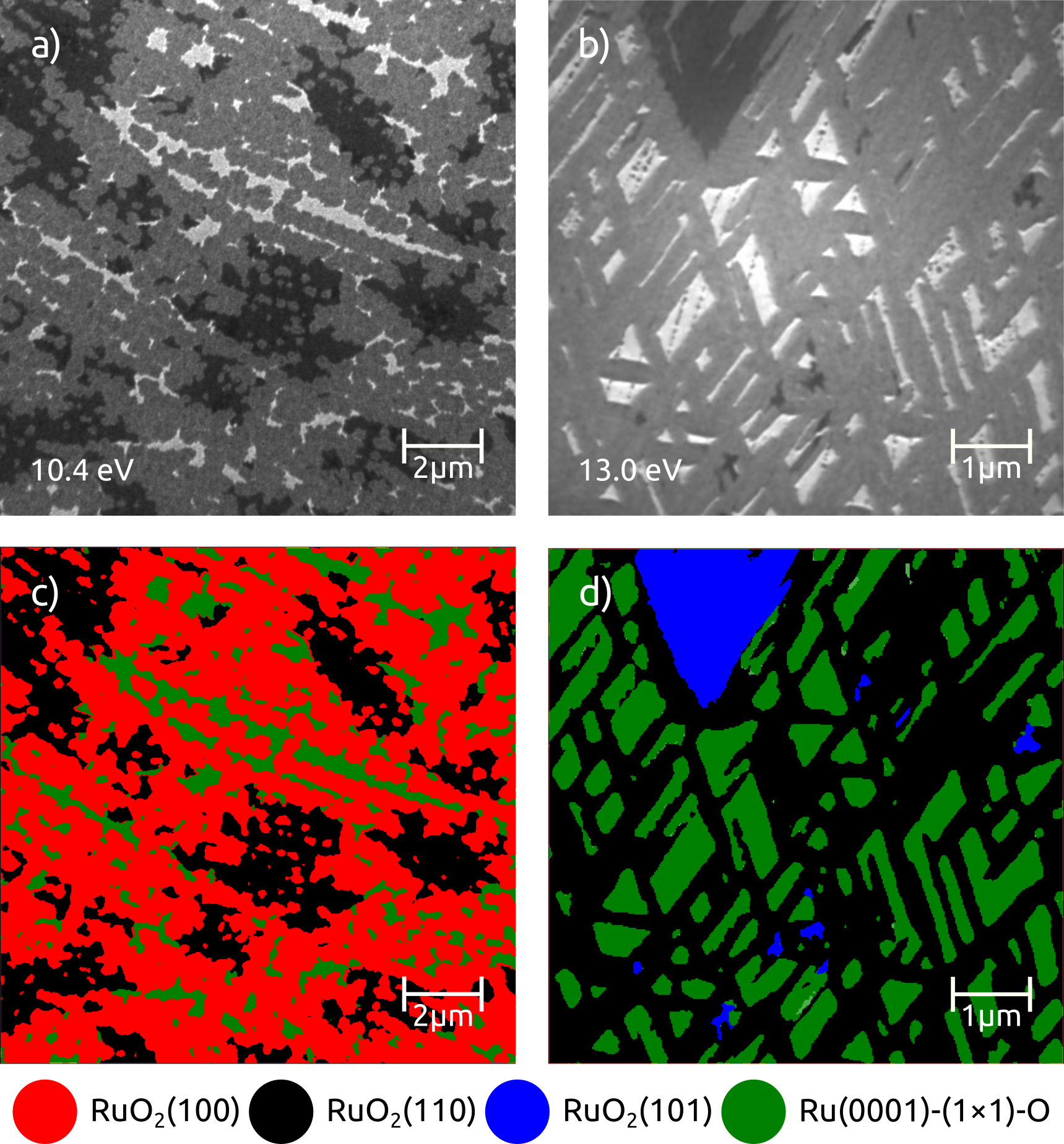}
		\caption{Two preparations showing the two characteristic types of \ce{RuO2} island distribution on Ru(0001). a) and b) show LEEM images of the surface. c) and d) depict the same regions with a color code showing the phases as determined from an I(V)~curve cluster analysis. Both samples are prepared by oxidizing the Ru crystal at \SI{460}{\celsius}, but with the sample shown in b) and d), in a first step the sample was first briefly oxidized at \SI{415}{\celsius} to form more \ce{RuO2}(110) nucleation centers.}
		\label{fig:ruo2-prep}
	\end{figure}
	
	The \ce{RuO2}/Ru(0001) templates were prepared as indicated in the previous section. Similar to the findings\cite{over2012,flege2015} for Ru(0001) oxidation using molecular \ce{O2}, three \ce{RuO2} island types of distinct orientation were observed in addition to a ``grainy'' phase consisting of nanoscale, diversely oriented \ce{RuO2} crystallites. The oxidation temperature has a large effect on the island distribution so that it is possible to favor one of the different \ce{RuO2} phases\cite{flege2016,flege2018}.
	
	Performing cluster analysis of the LEEM I(V) image stacks and comparing the resulting curves to experimental as well as calculated I(V) spectra from literature \cite{flege2015}, sample regions can be assigned to \ce{RuO2}(100), \ce{RuO2}(110), \ce{RuO2}(101), ``grainy'' \ce{RuO2} and a (1$\times$1)-O adlayer on the Ru(0001) as shown in Figure \ref{fig:ruo2-prep}. This assignment is verified by the locally obtained LEED patterns of each region (see also Figure \ref{fig:leed-110-100}). When not hindered by other structures, \ce{RuO2}(110) tends to form fully relaxed\cite{kim2001} elongated islands in three different rotational domains. Whereas \ce{RuO2}(100) islands also grow in three rotational domains, they are more compact. Because they  coalesce quickly, the individual domains are quite small and can often not be observed separately in {\si{\micro}LEED}. On some samples, also \ce{RuO2}(101) islands could be observed filling up areas between elongated \ce{RuO2}(110) islands. Some parts of the substrate Ru(0001) surface are left after oxidation; they are only covered by a (1$\times$1)-reconstructed oxygen adlayer.
	
	This surface layout allows for an in-depth investigation of \ce{VO2}/\ce{RuO2} growth on different orientations at the same time. With lateral dimensions in the micrometer range, the \ce{RuO2} islands are sufficiently large to serve as ``virtual substrates'' for \ce{VO2} film growth in the context of this work.
	
\subsection{Vanadium oxidation state}
	As laid out in the experimental section, in the next step the vanadium oxide films are prepared on the \ce{RuO2}/Ru islands by reactive MBE. 
	In order to evaluate whether in fact \ce{VO2} was grown, the exact stoichiometry of these films is determined. Through ex-situ XRR measurements of a characteristic sample, the film thickness is determined to be \SI{3.7}{\nano\meter} (see Supporting Information). Hence, in order to analyze the stoichiometry throughout the whole film, XAS-PEEM as well as XPEEM are employed because of their differing electron emission depths. From these methods, we obtain local x-ray absorption and x-ray photoelectron spectra.
	
	The position of the V~L$_3$ absorption edge as well as the V~$2p_{3/2}$ photoemission peak directly reflect the vanadium oxidation state because the V~2$p$ core level binding energy depends on the atoms' local chemical environment. Whereas the literature values on the photoemission peaks are relatively consistent with a \SI{14.0(3)}{eV} binding energy difference between the O~$1s$ and the V~$2p_{3/2}$ core level peaks for the V$^{4+}$ oxidation state \cite{mendialdua1995}, the position of the most prominent peak at the V~L$_3$ x-ray absorption spectroscopy edge is reported at \SI{517.4}{eV} in recent publications \cite{ruzmetov2007, aetukuri2013} but has also been placed at values in the range of \SIrange{516.6}{519.2}{eV}, probably due to different photon energy calibrations \cite{abbate1993,cressey1993,maganas2014,quackenbush2013}.
	
	\begin{figure}[bt]
		\centering
		\includegraphics[width=\linewidth]{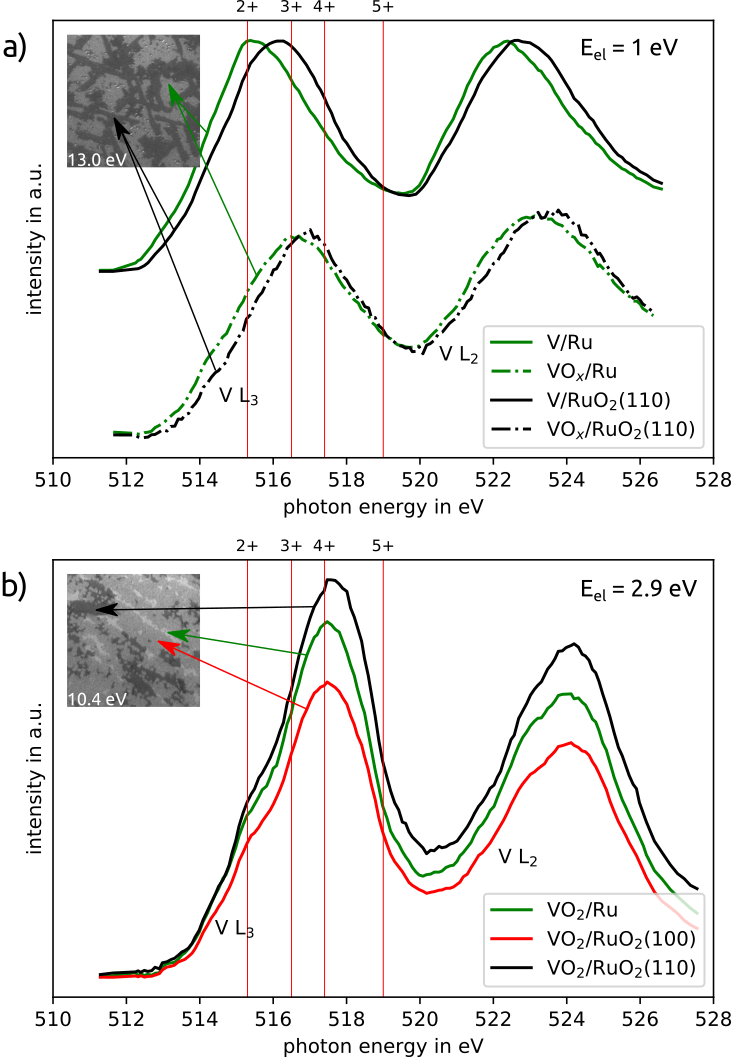}
		\caption{Local x-ray absorption spectra (XAS-PEEM) of the V~L edge. The vertical lines mark the positions of the main peak of the V~L$_3$ edge for different vanadium oxidation states (note that this does not apply for the observed shoulders; they do not imply partial reduction). Both spectra are taken with $\vec{E}\perp$ $c_R$ axis.
			a) After depositing just vanadium on the \ce{RuO2}/Ru template (solid lines) and after subsequent oxidation (dotted lines).
			b) After reactive MBE, providing vanadium and oxygen at the same time.
		}
		\label{fig:XAS}
	\end{figure}
	
	Figure \ref{fig:XAS}a shows x-ray absorption spectra of a sample where the \ce{VO2} film was grown cyclically by alternating between vanadium evaporation and subsequent oxidation inspired by the growth recipe of \citet{tashman2014}. Directly after depositing vanadium onto the \ce{RuO2}/Ru mesoscale islands, the main peak at the V~L$_3$ absorption edge is observed at \SI{516.1}{eV} for V/\ce{RuO2}/Ru regions which reflects a vanadium oxidation state between 2+ and 3+.
	On the Ru(0001) surface regions where there was only oxygen from the (1$\times$1)-O adlayer present before evaporating the vanadium, the vanadium is less oxidized (V~L$_3$ edge main peak at \SI{515.4}{eV}). This suggests that the vanadium takes oxygen from the \ce{RuO2} (thus reducing it) while the adlayer on the Ru surface provides a smaller oxygen reservoir.
	Subsequent exposure of this film to atomic oxygen obviously further oxidizes the vanadium, but it does not quite reach \ce{VO2} stoichiometry, neither on the \ce{RuO2} nor on the Ru substrate surface. This demonstrates that a cyclic growth recipe consisting of alternating vanadium evaporation and subsequent oxidation does not yield stoichiometric \ce{VO2} films even when using highly oxidizing atomic oxygen while it does lead to an oxygen depletion of the \ce{RuO2} islands during the initial vanadium deposition. Hence, in contrast to the \ce{VO2}/\ce{TiO2} system, this method is not viable for \ce{VO2}/\ce{RuO2}.
	
\begin{figure}[bt]
		\centering
		\includegraphics[width=\linewidth]{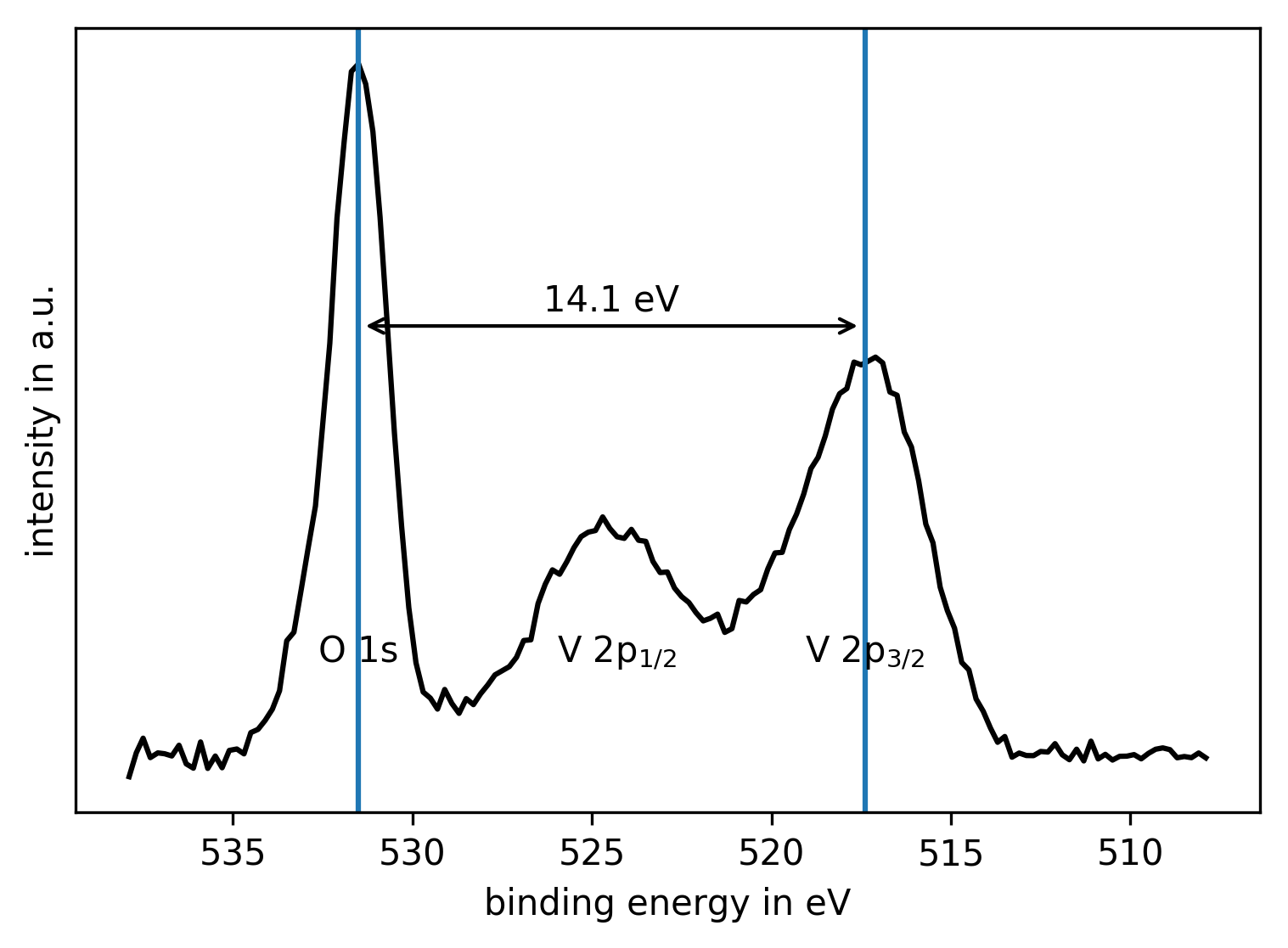}
		\caption{Local x-ray photoelectron spectrum of a \ce{VO2}/\ce{RuO2}(110) film  acquired by XPEEM (photon energy E$_\textrm{ph} = \SI{589.1}{\electronvolt}$). The electron energy difference of the marked photoemission peaks confirms the presence of stoichiometric \ce{VO2}.}
		\label{fig:XPS}
	\end{figure}

	On samples prepared instead by reactive MBE, i.e. simultaneous vanadium evaporation and oxidation, V$^{4+}$ forms directly. As can be seen in Figure \ref{fig:XAS}b, the position of the main peak maximum at the V~L$_3$ XAS edge is observed at \SI{517.4(1)}{eV}. This applies for both a growth temperature of \SI{200}{\celsius} followed by \SI{10}{\minute} of annealing at \SI{400}{\celsius} as well as for a growth temperature of \SI{400}{\celsius}. The energy observed for this peak is in accordance with previously reported values for \ce{VO2} with small deviations due to different instrument calibrations (see above). Also, the peak shape with the shoulder at about \SI{515.3}{\electronvolt} is in accordance with these studies \cite{aetukuri2013,abbate1993,ruzmetov2007,quackenbush2013} and the shoulder does \textit{not} imply a partial reduction of the film.
	XPEEM spectra (i.e. local XPS, Figure \ref{fig:XPS}) confirm this vanadium oxidation state with $\Delta \left[ E_\textrm{B}(\textrm{O}~1s), E_\textrm{B}(\textrm{V}~2p_{3/2}) \right] = \SI{14.1}{eV}$ on all present substrate phases.
	Due to the different kinetic energies and thus different emission depths of the detected electrons, XPS ($E_\textrm{el}$: \SIrange{50}{80}{eV}) is more surface sensitive \cite{hocker2015} than XAS ($E_\textrm{el}$: \SIrange{1}{3}{eV}). Therefore it can be concluded that the fourth vanadium oxidation state is found throughout the whole film of about \SI{3.7}{\nano\meter} on all substrate phases.
	
\subsection{Vanadium oxide surface structure}
	LEEM overview images recorded after the growth of VO$_2$ show the same morphology as before with the domain positions and sizes unchanged; the contrast however does change strongly as can be seen from comparing Figure \ref{fig:ruo2-prep} and the insets in Figure \ref{fig:XAS}. Therefore it is likely that the \ce{VO2} growth is specific to the individual underlying \ce{RuO2} and Ru phases. In order to investigate the structure of these differing \ce{VO2} films, local surface diffraction is the best suited tool. The \si{\micro}LEED mode of the LEEM allows for such a separate investigation of the structure on top of the distinct template phases.
	
	\subsubsection{(100)-oriented growth}	
	
	\begin{figure*}[bt]
		\centering
		\includegraphics[width=\linewidth]{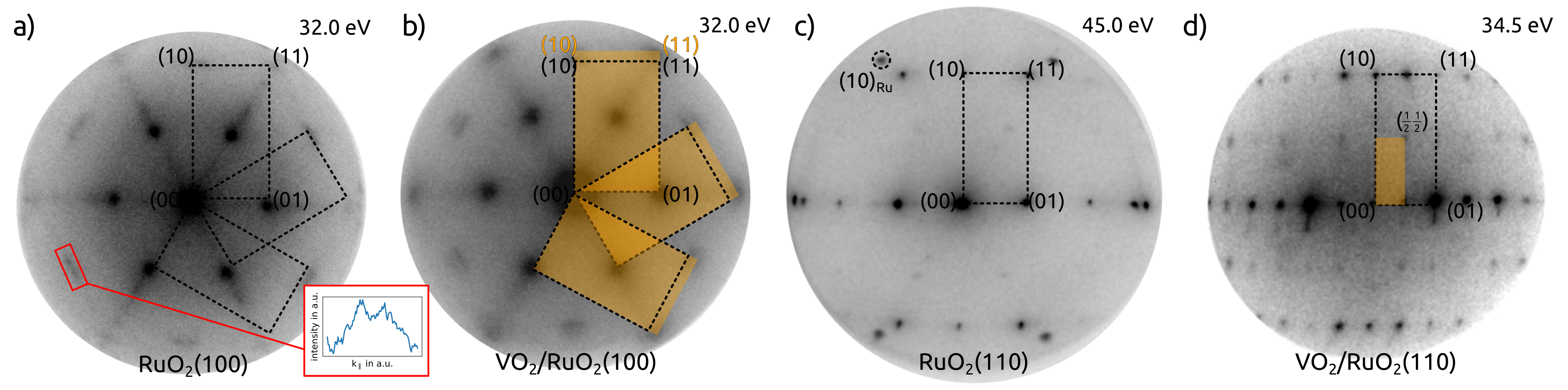}
		\caption{Local LEED images taken with an illumination spot of \SI{500}{\nano\meter} in diameter at \SI{200}{\celsius} sample temperature. 
			a) \ce{RuO2}(100) islands of three rotational domains. The unit cells are marked by dotted lines. A line profile along a (10) spot azimuthal elongation shows two maxima.
			b) Additional spots appear after \ce{VO2} growth on top of the \ce{RuO2}(100) islands. The resulting unit cells are marked in orange.
			c) Single domain \ce{RuO2}(110) pattern with the unit cell marked by a dotted line. 
			d) (2$\times$2)-reconstructed pattern after \ce{VO2} growth on top of the \ce{RuO2}(110). The resulting unit cell is marked in orange. Note that c) and d) are taken at different electron energies, changing the size of the Ewald sphere.}
		\label{fig:leed-110-100}
	\end{figure*}
	
	Like all orientations of epitaxial \ce{RuO2} islands on Ru(0001), (100)-oriented \ce{RuO2} grows in three rotational domains with rectangular unit cells. The domain size of this orientation is too small to get single domain LEED patterns with the smallest available illumination aperture, so the Ru(0001) substrate and all three rotational domains superimpose in the diffraction image shown in Figure \ref{fig:leed-110-100}a. Like in previous studies \cite{flege2016}, the (10) spots are azimuthally elongated with two maxima along that direction because of a misfit to the Ru(0001) that allows for two energetically equivalent registries. While the (11) spots are quite weak, the (01) spots seem broadened due to an overlapping reflection from a (2$\times$2)-3O-reconstructed oxygen adlayer next to the \ce{RuO2} islands that contributes to the diffraction image. This adlayer is a reduced variant of the Ru(0001)-(1$\times$1)-O adlayer. These facts lead to overall comparably weak diffraction spots of this island type.

	After \ce{VO2} film deposition at \SI{400}{\celsius} on these (100)-oriented islands, in the (10)-directions two radially separated reflections are visible (Figure \ref{fig:leed-110-100}b). The inner spots can be assigned to the previously observed \ce{RuO2}(100) while the outer spots are related to the \ce{VO2} grown on top. This confirms that the \ce{VO2} did indeed form crystalline films aligned to the \ce{RuO2} template on this phase. As the (10)-direction in the unit mesh of (100)-oriented \ce{RuO2} and \ce{VO2} represents the $c_R$ direction in the rutile unit cell, the ratio of (10) spot distances equates to the inverse of the ratio of lattice plane distances in this $c$ direction:
	\[
	\frac{d_{\textrm{RuO}_2(10)}}{d_{\textrm{VO}_2(10)}} = \num{1.075(5)}
	\]
	This matches fairly well with the ratio of the bulk lattice constants of rutile \ce{VO2} $\frac{c_{\textrm{RuO}_2}}{c_{\textrm{VO}_2,\textrm{R}}} = \num{1.088}$ \cite{rogers1993,rao1969}, confirming that the film grew almost fully relaxed. Along the $(1\overline{1}0)$ directions, the lattice mismatch is too small to make out the individual spots.
	The (11) diffraction spots almost coincide with the Ru(0001) substrate spots, making it seem like these spots of relaxed \ce{VO2} could originate from \ce{VO2} grown directly on the Ru(0001). However, this pattern does not occur further away from the \ce{RuO2}(100) islands on a pure Ru(0001) region.
	
	\subsubsection{(110)-oriented growth}
	
	For the (110)-oriented \ce{RuO2} islands, the rotational domains are much larger and it is therefore possible to obtain single-domain LEED patterns as shown in Figure \ref{fig:leed-110-100}c where, apart from very weak Ru(0001)-(2$\times$2)-3O-adlayer related spots and the substrate spots, only the reflections of the rectangular (110) unit mesh are visible. Measuring the ratio of distances from the specular reflection to the (10) and (01) spots, it is possible to calculate the ratio of the lattice constants in $c$ and $a$ direction: 
	\[
	\frac{c}{a} = \frac{\sqrt{2}\cdot d_{(01)}}{d_{(10)}} = \num{0.70(1)}
	\]
	Within the error margins, this matches the bulk value for \ce{RuO2}: $\frac{c_{\textrm{RuO}_2}}{a_{\textrm{RuO}_2}} = \num{0.692}$ \cite{rao1969}.
	
	Following \ce{VO2} growth, LEED patterns from these \ce{VO2}/\ce{RuO2}(110) regions reveal a very different picture compared to the one on the (100) orientation. Directly after deposition at \SI{200}{\celsius}, only the previously visible spots of the \ce{RuO2}(110) surface can be seen, having strongly diminished in intensity. During post-annealing the films in atomic oxygen however, their intensity increases again due to improved ordering. Then, starting at \SI{250}{\celsius} additional spots start to appear in the (01)-direction, corresponding to a (1$\times$2) reconstruction pattern. At about \SI{400}{\celsius}, this further evolves into the (2$\times$2) pattern shown in Figure \ref{fig:leed-110-100}d. The pattern clearly still originates from a rutile structure (or ``rutile-similar'' monoclinic) (110) surface while the sharp spots indicate a very good crystalline quality.
	
	Looking at the ratio of distances from the specular reflection to the (10) and (01) reflections, the ratio of the $c_R$ and $a_R$ lattice constants is calculated again:
	\[
	\frac{c}{a} = \frac{\sqrt{2}\cdot d_{(01)}}{d_{(10)}} = \num{0.69(1)}
	\]
	This clearly does not match the ratio of bulk-like \ce{VO2} ($\frac{c_{\textrm{VO}_2}}{a_{\textrm{VO}_2}} = \num{0.627}$ \cite{rogers1993}), but in fact has not changed significantly from the \ce{RuO2} bulk value. This was measured consistently for several frames of time-resolved LEED measurements during \ce{VO2} growth as well as during and after post-annealing. The variation in the absolute periodicities in the (10) and the (01) direction is only \SI{0.2}{\percent}.
	
	In the section on the vanadium oxidation state, it was confirmed that stoichiometric \ce{VO2} is present on all types of \ce{RuO2} islands. So the periodicity deduced from LEED still staying constant during growth means that a fully strained \ce{VO2} is now present on \ce{RuO2}(110). 
	Therefore, the vanadium dioxide also is the source of the (2$\times$2) reconstruction in Figure \ref{fig:leed-110-100}d.
	A (2$\times$2)-reconstructed \ce{VO2}(110) surface has been observed before in thin films on \ce{TiO2} substrates: LEED measurements by \citet{laverock2014} also show half spots in both crystallographic directions on the \ce{VO2}(110) surface for the insulating as well as for the conducting state, although they do only elaborate on a perceived (2$\times$1) superstructure that they attribute to the insulating (i.e. monoclinic) \ce{VO2} phase.
	
\begin{figure}[bt]
		\centering
		\includegraphics[width=0.5\linewidth]{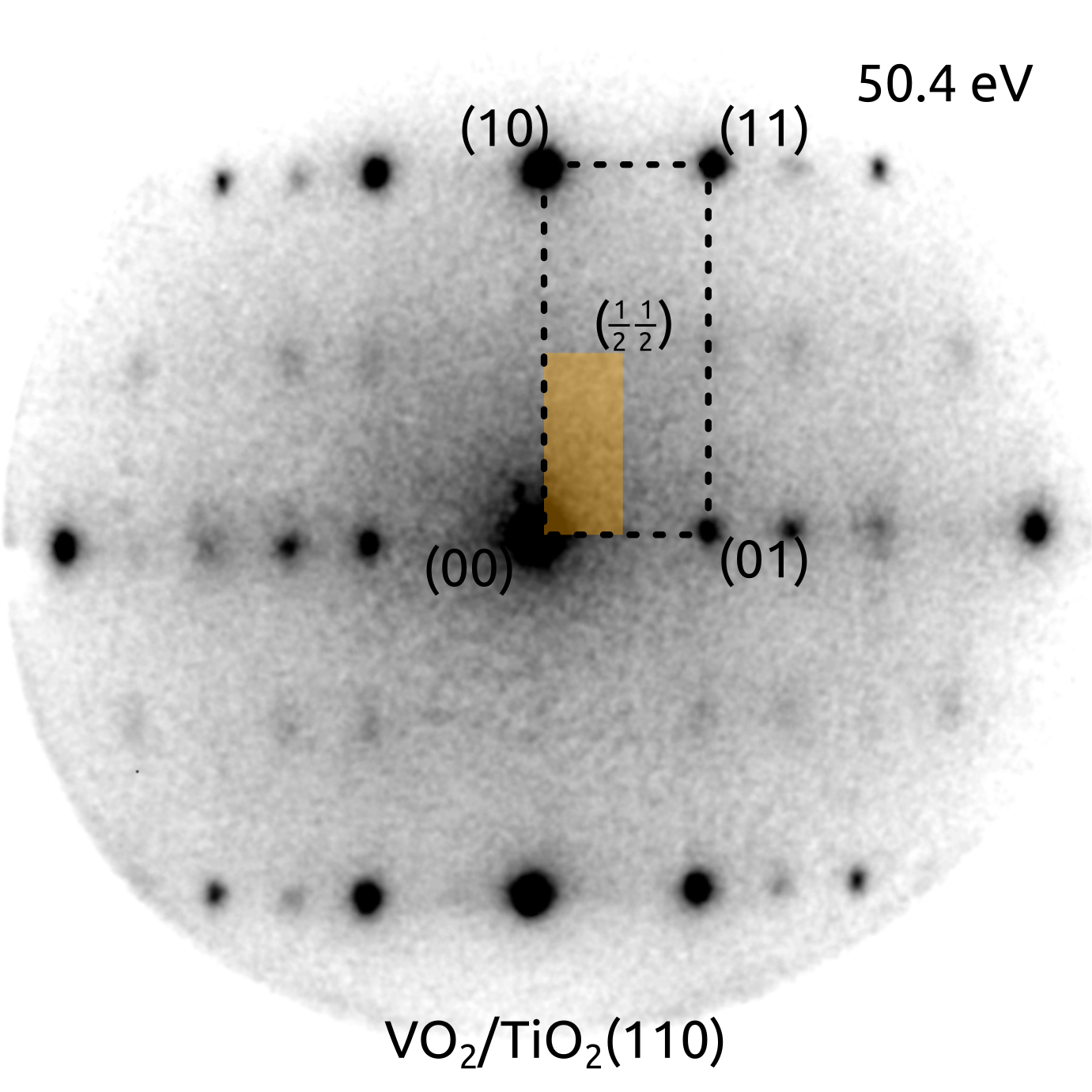}
		\caption{LEED image of a \ce{VO2}/\ce{TiO2}(110) thin film surface prepared by reactive MBE using atomic oxygen from a thermal cracker after shortly annealing to \SI{600}{\celsius} in atomic oxygen. The (2$\times$2) reconstructed unit mesh is marked in orange.}
		\label{fig:leed-tio2}
	\end{figure}

	We tried to reproduce this reconstruction by growing \SIrange{2}{4}{\nano\meter} \ce{VO2} thin films on \ce{TiO2}(110) with reactive MBE using the same recipe as for \ce{VO2}/\ce{RuO2}, i.e. deposition of vanadium while providing atomic oxygen simultaneously at \SI{200}{\celsius}. During annealing in atomic oxygen, a (2$\times$2) reconstruction indeed appears (see Figure \ref{fig:leed-tio2}). This reconstruction is stable while heating up to \SI{650}{\celsius}. 
	This observation contradicts the hypothesis that the surface reconstruction results from the bulk crystal distortions originating from the semiconductor-metal transition but in fact suggests that it is mainly determined by the occupation and geometry of oxygen surface sites as has been investigated by \citet{mellan2012}. Although in their study no (2$\times$2)-reconstructed surface is included, it is obvious that the oxygen chemical potential determines the occupation of surface sites and thus leads to the formation of different surface reconstructions. In comparison to previous \ce{VO2}(110) surface studies\cite{sambi1997,agnoli2004}, the oxygen chemical potential provided in this work is much higher owing to the use of a thermal cracker, making it plausible for a new type of surface reconstruction to appear. That this reconstruction appears also on \ce{VO2}/\ce{RuO2} is an indicator for the good quality of those films.
	While the surface structure remains virtually unaffected by the structural transition in the bulk, it might still be possible to determine an electronic switching of the film via very careful local, temperature-dependent XAS measurements \cite{aetukuri2013,quackenbush2016}. In the scope of this study however, such a behavior could not be detected. 
	
	Both the (100)- and the (110)-oriented surface have the $c_R$ axis lying in-plane. In the other in-plane direction, they also have an equal lattice mismatch. \citet{laverock2012} note that they achieved a much higher strain along the $c_R$ axis when growing \SI{40}{\nano\meter} thick \ce{VO2} films on \ce{TiO2}(100) (\SI{3.7}{\percent}) than on \ce{TiO2}(110) (\SI{1.7}{\percent}) substrates.	
	The higher actual strain achieved on the (100) orientation might be caused by the lower stability of the \ce{VO2}(100) surface, also making early relaxation like we observe it plausible. The (110) surface is, in fact, the most stable one\cite{mellan2012} and can, as shown above, withstand pseudomorphic growth on \ce{RuO2}.

\section{Conclusion}
	Using a combination of LEEM and synchrotron-based PEEM techniques that allow for a chemical as well as structural analysis, we characterized the growth of \SI{3}{\nano\meter} thick \ce{VO2} thin films on \ce{RuO2}/Ru(0001) substrates featuring \ce{RuO2} islands of different orientation simultaneously. These templated substrates were created in-situ by oxidizing a ruthenium single crystal by atomic oxygen.
	
	Locally obtained x-ray absorption spectra as well as x-ray photoelectron spectra of the vanadium oxide films prepared by reactive MBE are in good agreement with previous \ce{VO2} studies on other substrates, confirming the \ce{VO2} stoichiometry on \ce{RuO2}(110), \ce{RuO2}(100) as well as on the bare Ru(0001).
	
	On \ce{RuO2}(100), local electron diffraction revealed that the \ce{VO2} films are of good crystalline quality and the orientation aligns to the underlying \ce{RuO2}. However, it was found that these (100)-oriented films are fully relaxed, the surface unit mesh periodicity matching the bulk \ce{VO2} $c_R$ lattice constant.
	
	On \ce{RuO2}(110) on the other hand, the ratio of the in-plane lattice constant as well as their absolute values evidence pseudomorphic growth within the limits of accuracy of the measurements. Considering the lattice mismatch of \SI{8.78}{\percent} along the $c_R$ axis on \ce{RuO2}(110), this proves \ce{VO2} thin films to be extremely strainable and expands the list of suitable \ce{VO2} growth substrates. 
	Electron diffraction also shows that the \ce{VO2}(110) film surface reconstructs to a (2$\times$2) superstructure. In the present study, this reconstruction was reproduced for \ce{VO2}/\ce{TiO2}(110) and likely arises because of the high oxygen potential available during growth with atomic oxygen that leads to a previously undescribed oxygen termination.
	
	These findings open the window for further research in massively strained \ce{VO2} thin films and the possibility of decoupling or suppressing the Peierls transition in favor of the Mott transition of \ce{VO2}.

\begin{acknowledgement}
	The authors gratefully acknowledge Vedran Vonk and Andreas Stierle from the DESY in Hamburg for ex-situ XRR measurements.
	
	JOK and JIF acknowledge financial support from the Deutsche Forschungsgemeinschaft DFG under grant number 362536548.
	JOK further acknowledges support from DFG grant number 408002857 as well as by the Institutional Strategy of the University of Bremen, funded by the German Excellence Initiative. 
\end{acknowledgement}

\begin{suppinfo}
	Descriptions of the XRR ex-situ measurements used for determining the film thicknesses.
\end{suppinfo}

{
	\footnotesize
	\bibliography{references}
}

\newpage
~
\newpage

\begin{minipage}{\textwidth}
	\section{For Table of Contents Use Only}
	\bigskip
	\begin{center}
		\textbf{\Large Massively strained \ce{VO2} thin film growth on \ce{RuO2}}
		
		\medskip
		\begin{minipage}{0.7\textwidth}
			Simon Fischer, Jon-Olaf Krisponeit, Michael Foerster, Lucia Aballe, Jens Falta, and Jan Ingo Flege
		\end{minipage}
		
		\bigskip
		\includegraphics{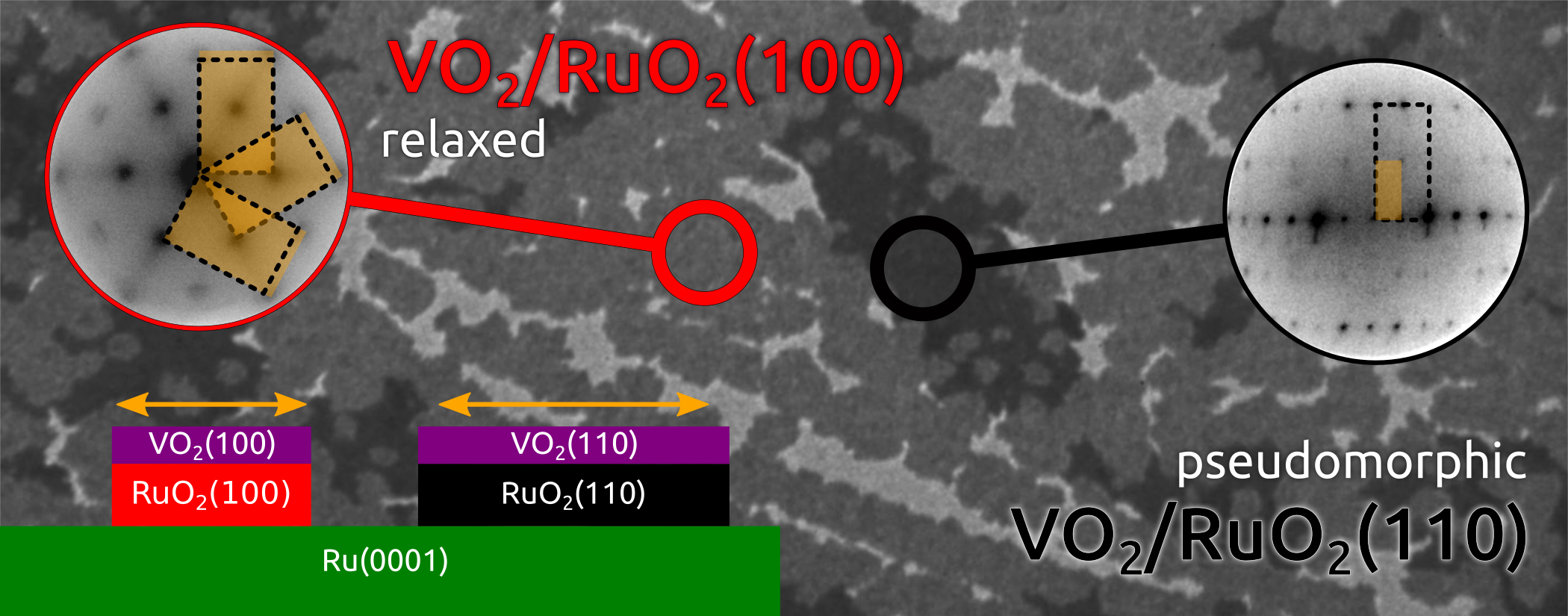}
	\end{center}

	\bigskip
	\emph{Synopsis:} Heavily strained \ce{VO2} films were grown on \ce{RuO2}(110) islands by atomic oxygen-assisted reactive MBE. These films exhibit a previously undescribed (2$\times$2) reconstruction. On \ce{RuO2}(100) islands, \ce{VO2} was found to grow immediately relaxed.
\end{minipage}

\end{document}